\documentclass[a4paper, 12pt]{article} 
\usepackage[T2A]{fontenc}              
\usepackage[english]{babel}
\usepackage{amssymb,amsmath,amsthm,amsfonts}
\usepackage[hmargin={0.8cm,0.8cm}, top=1cm]{geometry} 

\usepackage{graphicx} 
\usepackage{multirow}

\newtheorem{theorem}{Theorem}
\newtheorem{lemma}{Lemma}
\newtheorem{proposition}{Proposition}
\newtheorem{statement}{Statement}

\theoremstyle{definition}
	\newtheorem{definition}{Definition}
	
	\newtheorem{example}{Example}

\usepackage{hyperref} 
\hypersetup{colorlinks=true, linkcolor=blue, urlcolor=cyan,}

\numberwithin{equation}{section}

\author{A.V.~Kuliga, I.N.~Shnurnikov}

\title{Turnover of investment portfolio via covariance matrix of returns\footnote{This work was supported by the grant of the state program of the «Sirius» Federal Territory «Scientific and technological development of the «Sirius» Federal Territory» (Agreement №18-03 date 10.09.2024).
}.}
\date{}

\begin{document} 
		
	\maketitle
	
\vspace{-7mm}
\begin{center}
	\footnotesize{Sirius University of Science and Technology, \\
	Olimpiyskiy ave. b.1, Sirius, Krasnodar region, Russia, 354340.}
\end{center}

\abstract{An investment portfolio consists of $n$ algorithmic trading strategies, which generate vectors of positions in trading assets. Sign opposite trades (buy/sell) cross each other as strategies are combined in a portfolio. Then portfolio turnover becomes a non linear function of strategies turnover. It rises a problem of effective (quick and precise) portfolio turnover estimation. Kakushadze and Liew (2014) shows how to estimate turnover via covariance matrix of returns. We build a mathematical model for such estimations; prove a theorem which gives a necessary condition for model applicability; suggest new turnover estimations; check numerically the preciseness of turnover estimations for algorithmic strategies on USA equity market.}


	
\section{Introduction}	
	Our work deals with turnover of an investment portfolio, which is build of $n$ algorithmic trading strategies. Worked out model may be applied to various strategies, including hedge funds ones. Algorithmic trading in hedge funds is thoroughly discussed in books of Chan \cite{Chan}, Tulchinsky et al. \cite{Tulch}. One may find examples of strategies and ideas for their research in the Kakushadze and Serur's book \cite{Zura_11}, Kakushadze's article \cite{Zura_10} and in the section \ref{sec:experiments}.
    
    We investigate the following practical problem: how to estimate portfolio turnover while sign opposite trades cross, see example \ref{ex:crossing} below. There are such estimations (Kakushadze and Liew \cite{Zura_1}, Kakushadze \cite{Zura_2, Zura_4}) via weights and turnovers of strategies and empirical covariance matrix of their returns. This matrix is commonly used e.g. for determining portfolio quadratical risk and for portfolio optimization. We built the mathematical model for the turnover estimates and find the necessary condition of their correctness.

    The remaining part of the article is organised as following. In the introduction below we discuss hedge funds, define crossing of trades, set a task and show its applicability, enumerate known turnover estimates and results of the paper. In the section \ref{sec:math} we build the mathematical model for portfolio turnover, formulate and prove theoretical results, claim practical estimations for portfolio turnover. In the section \ref{sec:experiments} we describe experiments, define numerical characteristics of trading strategies and <<metrics>> for turnover estimates. We conclude with the section \ref{sec:conclusions} discussing theoretical and practical applications of the work. 
    
	\subsection{Hedge fund industry and definitions.}
	\begin{definition} 
		A hedge fund uses complex methods of liquid assets management and risk control for achieving good returns/risk values and for protection against market risk.
	\end{definition}

    E.g. long and short positions on the equity market hedge each other, as well as assets and options on them do. First hedge fund was founded by A.~W.~Jones in 1949. There are\footnote{by January 2024, https://investingintheweb.com/blog/largest-hedge-funds-aum/} about 15000 hedge funds in the world, having in common about $\$ 4.5$ trillions assets under management (AUM). Biggest hedge funds allocate tens of billions dollars for algorithmic trading strategies\footnote{E.g. Renaissance Technologies, Two Sigma Investments and	D.~E.~Shaw have about $\$106$, $\$67.4$ and $\$45.7$ billions AUM correspondingly. Man Group (30.06.2024), Millennium Management and Citadel have about $\$178.2$, $\$57.6$ and $\$51.5$ billions AUM correspondingly, but allocate part of it for algorithmic trading strategies.}.
    One may build algorithmic strategies on equity and cryptocurrencies markets, options and futures. As we know, algorithmic trading hedge funds have the following divisions.
    
	\begin{itemize}
	\item Quantitative researchers, from couples to several hundreds of ones, seek out and test algorithmic trading strategies based on market inefficiencies. 
	
	\item Portfolio managers build portfolios from strategies, using optimization methods. One may use returns/risk ratios (Markowitz type), risk budgeting method and so on\footnote{there are plenty of literature devoted to portfolio optimization, let's  refer to \cite{Markowitz, Bruder, Zura_9} only.}.
	
	\item Head division allocates capital among portfolio managers. Here are used models with transaction costs and market impact as well as discretionary decisions.
	
	\item Execution division may use non trivial mathematical modelling for finding better market execution.
	
	\item Other divisions (data collection and processing and so on).
	\end{itemize}

    \begin{definition} 
        \textit{Algorithmic trading strategy} (known in hedge funds and referred to below as \textit{alpha}) is a program, which generates vectors of assets positions to be taken by predefined moments of time.
    \end{definition}

    \begin{definition} 
        \textit{Portfolio} is a linear combination of $n$ alphas, the coefficients of which we will call \textit{weights}.
    \end{definition}

    \begin{definition} 
        Moment \textit{turnover} of an alpha (or portfolio) is $l_1$ norm of difference of position vectors, which are calculated at consecutive time moments. In other words, moment turnover is a sum over all assets absolute values of position differences on the asset\footnote{see subsection \ref{subsec:alpha_stats} for turnover formula.}. 
        Alpha and portfolio moment turnover is not constant in time. Later on we will consider time--averaged moment turnover and refer to it simply as turnover.\footnote{in practice alpha (or portfolio) turnover is time--averaged moment turnover, divided by average alpha (or portfolio) capital. We assume that alpha capital is constant, portfolio capital is constant as there is no netting. Then we may omit dividing by capital for simplification.} 
    \end{definition}

    \begin{definition} 
        \textit{Return} of the $i$--th alpha we will consider as a random variable $\alpha_i$, for which we have samples based on historical data, and a new value is added after each time interval.
    \end{definition}

    These definitions are quite general, enough for our work applications. In practice there are additional restrictions on time moments, delay between calculations of positions and their realization, properties of position vectors and market execution procedure. E.g. one use dollar neutrality condition, which means that sum of positions for all assets is 0. Alpha numerical characteristics may be found in subsection \ref{subsec:alpha_stats}.

	\subsection{Crossing of trades.}

	Key ingredient is the following assumption on the alpha execution. \textit{Trades} of opposite signs (buys/sells) of an asset cross each other when alphas are combined in a portfolio. Crossing of trades reduce transaction costs, market impact and so on.
    		
    \begin{example} \label{ex:crossing}
	Let us take 4 tickers: SBER, VTBR, TCSG and POSI. Let yesterday money (e.g. in thousands roubles) positions of alphas 1 and 2 be (0, 500, -200, -300) and (250, -400, 250, -100) correspondingly, where negative number means short position on an asset. So alpha 1 shorts some papers of POSI on 300\footnote{we neglect the fact that number of papers is integer and we could only approximately achieve required positions} th.r. Let today money positions of alphas be (100, 100, 300, -500) and (200, -300, 300, -200) correspondingly. To achieve today positions alpha 1 needs to buy SBER for 100 th.r., sell VTBR for 400 th.r., buy TCSG for 500 th.r. and sell POSI for 200 th.r. Alpha 2 needs to sell SBER for 50 th.r., buy VTBR for 100 th.r., buy TCSG for 50 th.r. and sell POSI for 100 th.r. So daily turnover of alphas will be 1200 and 300 th.r. correspondingly. A portfolio is a sum of alphas 1 and 2, weights are equal to 1. If alphas are executed separately, then total exchange orders will be of 1500 th.r. While execution with crossing of trades will be of 1200 th.r., needing to buy SBER for 50 th.r., sell VTBR for 300 th.r., buy TCSG for 550 th.r. sell POSI for 300 th.r. So crossing of trades reduces exchange orders from 1500 to 1200 th.r. and hence reduce transaction costs and so on.
	\end{example}
	
	Let us note that sign opposite \textit{positions} do not reduce each other (reducing of positions is commonly called \textit{netting}). This means that every alpha is provided with some capital and it doesn't change until the next portfolio rebalance. 
		

	\subsection{Setting and relevance of the problem.}
		
	 Let us point out two facts.
		
	\begin{enumerate}
		\item Due to crossing of sign opposite trades portfolio turnover is less\footnote{theoretically <<not more than>>, but always less in practice.} than linear combination of alpha turnovers with the same weights. 
        		
		\item Exact calculation of turnover is possible but too long\footnote{for hundreds or thousands of assets for crypto or equity markets correspondingly.} for building portfolios via optimization methods.
	\end{enumerate}
	
	\textbf{Practical problem.}
		\textit{Find an approximation (estimation) for portfolio turnover in terms of alphas weights and turnovers and any other available alpha data.} 

    E.g. sample covariance matrix of returns was used in articles \cite{Zura_1,Zura_2, Zura_4} as additional data. Alpha weights usually are the solutions of an optimization problem (e.g. returns maximization for fixed risk). Turnover estimations could be applied in hedge funds, where the following conditions hold.
    
	\begin{enumerate}
		\item Portfolio is built of $n$ alphas, which take positions on $s$ assets.
		\item Sign opposite trades of an asset cross each other when corresponding alphas are combined in a portfolio.

		\item Portfolio returns take into account transaction costs, market impact and so on, which depend on portfolio turnover.
        
		\item Alpha weights calculation uses portfolio returns (e.g. as optimized function).
	\end{enumerate}
    
  Portfolio turnover estimations are used in \cite{Zura_3,Zura_5,Zura_6}. These articles show how to build optimal portfolios of huge number of alphas (hundreds of thousands). Sample covariance matrix becomes generate because its rank is not more than the number of observations, which is of order thousands usually. So one replace the covariance matrix by <<similar>> non generate one, as in \cite{shrink_1} or with risk factors for alphas as in \cite {Zura_3,Zura_5,Zura_6}.
  
In \cite {Zura_3} one maximizes portfolio Sharpe ratio\footnote{formula may be found in subsection \ref{subsec:alpha_stats}} with linear costs and non linear market impact. There was built an algorithm, which give an exact solution by finite time.

In \cite{Zura_5} one maximizes portfolio returns under a constraint on the Sharpe ratio. By using a factor model for the covariance matrix the original optimization problem was reduced to a finite set of problems of finding zeroes of the line defined functions.

In \cite{Zura_6} were obtained alpha risk factors. Optimization of the Sharpe ratio for small deformed sample covariance matrix is reduced to the problem of weighted regression with respect to principal components.

	\subsection{Empirical estimations of portfolio turnover.}
	
	Articles \cite{Zura_1,Zura_2,Zura_4} are key to our work, as they investigate the portfolio turnover provided crossing of sign opposite trades of different alphas. Let us introduce the following notations.
	\begin{itemize}
		\item $x_i$ and $\tau_i$ --- weight and turnover of the $i$-th alpha, where $i = 1, \dots, n$.
		\item $\Psi$ --- $n \times n$ covariance matrix of alpha returns.
		\item $\psi^{(p)}$ and $V^{(p)} = \left(V_1^{(p)}, \dots,V_n^{(p)}\right)$ --- eigenvalues and orthonormal eigenvectors of the matrix $\Psi$ correspondingly.
		\item $\rho = \Psi_{12}$ --- correlation of two alphas returns. 
	\end{itemize}
		
	\textbf{Kakushadze and Liew, \cite[(4)]{Zura_1}.} Estimation for two alphas portfolio turnover, $x_1 > 0, x_2 > 0$:
	\begin{equation} \label{eq:zura_est_2}
	T_{\ast} = \frac{1+\rho}2(\tau_1 x_1 + \tau_2 x_2) + \frac{1-\rho}2|\tau_1 x_1 - \tau_2 x_2|.
 	\end{equation}
 	Also in \cite{Zura_1} a turnover limit for $n \to \infty$ was found by multiple using of (\ref{eq:zura_est_2}). If all alpha weights and turnovers and pairwise correlations of returns are equal to each other and are equal to $\frac 1n$, $\tau$ and $\rho$
    correspondingly, then the portfolio turnover limit is equal to $\tau\rho$.
    
 	\textbf{Kakushadze, \cite[(25)]{Zura_2}.} Turnover estimation for $n$ alphas portfolio.
 	\begin{equation} \label{eq:zura_est_spectral}		
 	T_{\ast} = \frac{1}{\sqrt{n}}\sum_{p=1}^n\psi^{(p)}\left|\sum_{i=1}^nV_i^{(p)}\tau_i|x_i|\right|.
   	\end{equation}	
    Let us note, that (\ref{eq:zura_est_2}) is a special case of the estimation (\ref{eq:zura_est_spectral}). Also in \cite{Zura_2} approximate formulas \cite[(31),(33),(34)]{Zura_2} were obtained for portfolio turnover provided $n \to \infty$ and some conditions on alpha weights and turnovers and covariance matrix of returns. In \cite{Zura_4} Kakushadze suggests that alpha returns contain $F$ risk factors. Then covariance matrix of returns could be approximated using covariance matrix of risk factors, relationships between risk factors, alpha returns and specific alpha risks.
    Then one use \cite{Zura_2} formulas for approximated matrix to deduce asymptotic estimations for portfolio turnover.
  	

	\subsection{New results.}
	
	\begin{itemize}
		\item We build a mathematical model for portfolio turnover estimations which depend on the covariance matrix of alpha returns, see section \ref{subsec:math_model}. 
		
		\item Main theorem \ref{th:main} is proved. 
		
		\item We obtain a correctness condition \ref{st:correctness} for applying portfolio turnover estimations, which depend on the covariance matrix of alpha returns.

		\item New portfolio turnover estimations are proposed, see formulas (\ref{new_est}). 
		
		\item Numerical experiments are carried out to check exactness and compare turnover estimations, see section \ref{sec:experiments}. 
		
	\end{itemize}
		

\section{Portfolio turnover model.} \label{sec:math}
	\subsection{Mathematical model.}	\label{subsec:math_model}
	
	\begin{definition}
	Let us call $n$--dimensional ($n \geq 2$) random vector $\alpha = (\alpha_1, \dots, \alpha_n)$ on the probability space $(\Omega, \mathcal{F}, \mathrm{P})$ \textit{admissible}, it the variance $\mathrm{D}\alpha_i = 1$ for all $i = 1, \dots, n$ and non-zero linear combination of vector components cannot be equal to a constant almost surely.
	\end{definition}

    We may assume that the returns of $n$ alphas form an admissible vector. Indeed, one may achieve $\mathrm{D}\alpha_i = 1$ by dividing the position vector of the $i$-th alpha by $\sqrt{\mathrm{D}\alpha_i}$. For dollar-neutral alphas with delay 1 the return (holding pnl) is equal to the scalar product of the position vector and the vector of asset returns for the previous trading interval, see formula (\ref{eq:holding_pnl}).

    Linear independence with a constant means that the alpha returns are linear independent and their linear combinations cannot yield non-zero risk-free profit (it follows from market no-arbitrage axiom).

	\begin{definition}
		By $V(\alpha)$ we denote the linear space of linear combinations $\sum_{i=1}^n x_i\alpha_i,$ where  $x_i \in \mathbb{R}$ and $\alpha = (\alpha_1, \dots, \alpha_n)$ is an admissible vector.
	\end{definition}

	For a fixed set of alphas the space $V(\alpha)$ is the space of returns of all possible portfolios. The set of numbers $(x_1,\dots, x_n)$ uniquely determines both the portfolio and the random variable from $V(\alpha)$. Functions on portfolios, such as turnover, can be considered as functions on the space $V(\alpha)$.
	
	\begin{definition}
	A function $f: V \to \mathbb{R}$ defined on the linear space $V$ is called \textit{absolutely homogeneous} of degree 1 if $f(\lambda v) = |\lambda|f(v)$ for all $\lambda \in \mathbb{R}$ and $v \in V$. 
	\end{definition}
	
	When a portfolio is multiplied by a constant $\lambda$, its turnover is multiplied by $|\lambda|$. This means that turnover is an absolutely homogeneous function on $V(\alpha)$. Let us introduce the following notations.

	\begin{itemize}
		\item A portfolio $\mathcal{P} = \mathcal{P}(x,\mathcal{A}) = \sum_{i=1}^n x_i\mathcal{A}_i$ is defined  by the set of $n$ weights $x=(x_1, \dots, x_n)$ and the set of $n$ alphas 
		$\mathcal{A} = (\mathcal{A}_1, \dots, \mathcal{A}_n)$. 
		
		\item Let $\tau(\mathcal{P})$ and $\tau(\mathcal{A}) = (\tau(\mathcal{A}_1), \dots, \tau(\mathcal{A}_n))$ be the turnovers of the portfolio and the set of alphas correspondingly.
		
		\item Let $\alpha(\mathcal{A})$ and $C(\alpha(\mathcal{A}))$ be a random vector of alphas returns and its covariance matrix correspondingly.
		
		\item Let $\mathbb{A}$ be a finite dimensional vector space including $\mathcal{A}_1, \dots, \mathcal{A}_n.$
		
	\end{itemize} 	
	
	\textbf{Main assumption of the model.}
		\textit{The portfolio turnover is expressed by some function of the alpha weights, turnovers and covariance matrix of returns. That is
		\begin{equation}\label{eq:turnover_assumption}
			\tau(\mathcal{P}(x,\mathcal{A})) = g(x,C(\alpha(\mathcal{A})),\tau(\mathcal{A}))		
		\end{equation}
		for all $x_i \in \mathbb{R}$ and $\mathcal{A}_i \in \mathbb{A}.$
	}

	\subsection{Theoretical results.}
    The following theorem describes absolutely homogeneous functions of degree 1, for which values of the sum depend only on the covariance matrix and the values of the terms. It turns out that 
    all such functions $f$ are equal to the standard deviation up to a multiplicative constant. Recall that $C$ and $\mathrm{D}$ denote the covariance matrix of several and the variance of one random variable correspondingly. 

	\begin{theorem}\label{th:main}
		Let $\alpha$ be an admissible random vector and $f: V(\alpha) \to \mathbb{R}$ be an absolutely homogeneous function of degree 1. Then the existence of a function $F$ such that
		\begin{equation} \label{eq:theorem_1}
			f(\xi_1 + \xi_2)= F(C(\xi_1,\xi_2), f(\xi_1), f(\xi_2))  \quad \text{for all} \quad \xi_1, \xi_2 \in V(\alpha),
		\end{equation}
		is equivalent to the existence of a constant $f_0$ such that 
		$f(\xi) = f_0\sqrt{\mathrm{D}(\xi)}$ for all $\xi \in V(\alpha).$ 
	\end{theorem}
	
	Let us apply the model assumption (\ref{eq:turnover_assumption}) for weights 
	$\bar{x}=(1,1,0,\dots,0)$ and alpha set $\mathcal{\bar{A}} = (\mathcal{A}_1, \mathcal{A}_2, 0, \dots, 0)$. Then for all $\mathcal{A}_1, \mathcal{A}_2 \in \mathbb{A}$:
	$$
	\tau(\mathcal{A}_1 + \mathcal{A}_2) = g(\bar{x},C(\alpha(\mathcal{\bar{A}})),\tau(\mathcal{\bar{A}}))
	=\bar{g}(C(\alpha(\mathcal{A}_1), \alpha(\mathcal{A}_2)), \tau(\mathcal{A}_1), \tau(\mathcal{A}_2)).
	$$	
	By theorem \ref{th:main} there exists a constant $f_0$ such that $\tau(\mathcal{A}_i) = f_0\sqrt{\mathrm{D}(\alpha(\mathcal{A}_i))}$ for all $i = 1, \dots, n.$ So, we have proved the following.

	\begin{statement} \textbf{Model correctness condition.} \label{st:correctness}
	If the turnover of a $n$ alpha portfolio is expressed as a function of the covariance matrix, weights and turnovers of alphas, then the ratio of turnover to the standard deviation of alpha returns should be the same for all alphas.
	\end{statement}

    From theorem \ref{th:main} we also deduce the following.
	
    \begin{statement} \textbf{Theoretical turnover estimation.} \label{st:theoretical_est}
    Let the model assumption (\ref{eq:turnover_assumption}) be true, $\tau_i$ be the turnover of $i$-th alpha for $i = 1, \dots, n$, $C$ be the covariance matrix of returns. By $\kappa$ we denote the ratio $\frac{\tau_i}{\sqrt{C_{ii}}}$ (it doesn't depend on $i$).  Then the turnover of the portfolio with weights $x=(x_1,\dots,x_n)^{\top}$ is equal to
		\begin{equation} \label{eq:theoretical_est}
			T_{\ast} = \kappa \sqrt{x^{\top}Cx}.
		\end{equation}
	\end{statement}

    The formula (\ref{eq:theoretical_est}) gives a different result compare to Kakushadze and Liew estimation \cite[(19)]{Zura_1} for $\tau_i = \kappa$ and covariance matrix
$$
 C_{ij}(\alpha) = 
 \begin{cases}
 	\rho, & i \neq j \\
 	1, & i=j
 \end{cases}\quad \text{for all} \quad i,j \in \{1,\dots,n\}.
 $$
Indeed, in \cite[(19)]{Zura_1} estimation is $T_{\ast} = \kappa(\rho + \frac{1 - \rho}n)$, whereas by formula (\ref{eq:theoretical_est}) estimation is $T_{\ast} = \kappa\sqrt{\rho + \frac{1 - \rho}n}$. The difference of the results could be explained by the fact that in \cite{Zura_1} the turnover estimation (\ref{eq:zura_est_2}) used many times, while it is incorrect due to the proposition \ref{prop:zura_2_uncorr}.

\begin{proposition} \label{prop:zura_2_uncorr}
    Let $\alpha$ be an admissible random vector and $f: V(\alpha) \to \mathbb{R}$ be an absolutely homogeneous function which is not zero constantly. Then there exist random variables $\xi_1, \xi_2 \in V(\alpha)$ such that
	$$
	f(\xi_1 + \xi_2) \neq \frac{1+\rho(\xi_1, \xi_2)}2(f(\xi_1) + f(\xi_2)) + \frac{1-\rho(\xi_1, \xi_2)}2|f(\xi_1)- f(\xi_2)|.
	$$
\end{proposition}


    \subsection{Practical turnover estimations.}
    In practice the ratio of turnover $\tau_{i}$ to the standard deviation of returns $std_{i} = \sqrt{C_{ii}}$ may differ by several times\footnote{e.g. for dollar neutral alphas on USA equity market. Another evidence that this ratio isn't constant may be found in the article by Kakushadze and Tulchinsky \cite[Fig. 4]{Zura_12}.} among $n$ alphas. Therefore, for $\kappa$ from the formula (\ref{eq:theoretical_est}) one can choose some <<average>> of them. In the estimations $T_{*1}$ and $T_{*2}$ we use the arithmetic and geometric means correspondingly, in the estimation $T_{*3}$ --- the weighted average with weights $x$. Let us denote by $\sigma = \sqrt{x^{\top}Cx}$ the standard deviation of the portfolio return, where alpha weights are $x$.

\begin{multline} \label{new_est}
	\hfill
	T_{*1} = \frac 1n\sum_{i=1}^n\frac{\tau_{i}}{std_{i}}\sigma,  
	\hfill
	T_{*2} = \left(\frac{\tau_{1}\cdot\ldots\cdot\tau_n}{std_{1}\cdot\ldots\cdot 	std_n}\right)^\frac{1}{n}\sigma,	
	\hfill
	\\
	\hfill
	T_{*3} = \frac{\sum_{i=1}^n\frac{x_i\tau_{i}}{std_{i}}}{\sum_{i=1}^n x_i}\sigma, 	
	\hfill
	T_{*4} = \frac{\sum_{i=1}^n\tau_{i}}{\sum_{i=1}^n std_{i}}\sigma.
	\hfill
\end{multline}


\subsection{Proofs.}	

\begin{lemma} \label{lem:1}
    For every admissible random vector $\alpha$ there exists an admissible random vector $\bar{\alpha}$ such that $V(\bar{\alpha}) = V(\alpha)$ and the matrix $C(\bar{\alpha})$ is identity.
\end{lemma}

    If $\alpha = (\alpha_1, \dots, \alpha_n)^\top$ is a random vector written as a column, and $A$ is a $n \times n$ matrix, then $C(A\alpha) = A C(\alpha)A^\top$. Indeed,
\begin{multline*}
	C(A\alpha) = \mathrm{M} (A\alpha - \mathrm{M} A\alpha)((A\alpha)^\top - \mathrm{M}(A\alpha)^\top) = \\
	= \mathrm{M} A (\alpha - \mathrm{M}\alpha)(\alpha^\top - \mathrm{M}\alpha^\top)A^\top = A C(\alpha)A^\top\\
\end{multline*}

\noindent
A symmetric matrix $C(\alpha)$can be reduced to the principal axes, i.e. find an orthogonal matrix $H$ such that $C(H\alpha) = HC(\alpha)H^\top = B$, where $B$ is a diagonal matrix. Let $\bar{\alpha} = B^{-\frac 12}H\alpha$, then $C(\bar{\alpha})$ is a identity matrix, $\bar{\alpha}$ is an admissible random vector and $V(\alpha) = V(\bar{\alpha})$. The non-degeneracy of the matrix $B$ (and the existence of $B^{-\frac 12}$) follows from the fact that the components of $\alpha$ are linearly independent with a constant. 
$\hfill \square$


\medskip
\textbf{The proof of proposition \ref{prop:zura_2_uncorr}.} 

By lemma \ref{lem:1} we may assume that the matrix $C(\alpha)$ is identity. We can also assume that $f(\alpha_1) \geq f(\alpha_2)$ and $f(\alpha_1) \neq 0.$ Indeed, since $f \not\equiv 0$ there exists $\beta_1 \in V(\alpha)$ such that $f(\beta_1) \neq 0$ and $\mathrm{D}\beta_1 = 1$. We extend $\beta_1$ to an admissible random vector $\beta = (\beta_1, \dots, \beta_n)$ with the identity matrix $C(\beta)$, $V(\beta) = V(\alpha)$ and assume that the vector $\alpha$ was initially equal to the vector $\beta$. The inequality $f(\alpha_1) \geq f(\alpha_2)$ could be achieved by rearranging the components of the vector $\alpha$.

Let us assume the contrary, i.e. that for all $\xi_1,\xi_2 \in V(\alpha)$ 
\begin{equation} \label{eq:sum_two_zura_local}
	f(\xi_1 + \xi_2) = \frac{1+\rho(\xi_1, \xi_2)}2(f(\xi_1) + f(\xi_2)) + \frac{1-\rho(\xi_1, \xi_2)}2|f(\xi_1)- f(\xi_2)|.
\end{equation}
Let us use (\ref{eq:sum_two_zura_local}) for $\xi_1 = \alpha_1,\ \xi_2 = \alpha_2$ and for $\xi_1 = \alpha_1,\ \xi_2 = -\alpha_2$:
\begin{equation*}
	f(\alpha_1 + \alpha_2) = \frac{f(\alpha_1) + f(\alpha_2)}2 + \frac{|f(\alpha_1) - 		f(\alpha_2)|}2 = f(\alpha_1 - \alpha_2) = \max\{f(\alpha_1), f(\alpha_2)\}
\end{equation*}
Let us use (\ref{eq:sum_two_zura_local}) for $\xi_1 = \alpha_1 + \alpha_2, \ \xi_2 = \alpha_1 - \alpha_2$, taking into account that $f(\alpha_1 + \alpha_2) = f(\alpha_1 - \alpha_2)$ and $\rho(\alpha_1 + \alpha_2, \alpha_1 - \alpha_2) = 0$:
\begin{equation*}
	f(2\alpha_1) = f((\alpha_1 + \alpha_2) + (\alpha_1 - \alpha_2)) = \frac{f(\alpha_1 + 	\alpha_2) + f(\alpha_1 - \alpha_2)}2 = \max\{f(\alpha_1), f(\alpha_2)\}
\end{equation*}
So, $2f(\alpha_1) = f(2\alpha_1) = \max\{f(\alpha_1), f(\alpha_2)\} = f(\alpha_1)$, hence $f(\alpha_1) = 0$, contradiction. $\hfill \square$


\medskip
\textbf{The proof of the theorem \ref{th:main}.}

\textit{Step 0.} From the existence of a constant $f_0$ it easily follows the existence of function $F$. Indeed,
\begin{equation*}  
	f(\xi_1 + \xi_2)= f(\xi_1)\sqrt{\frac{C_{11}(\xi_1,\xi_2) + 2C_{12}(\xi_1,\xi_2) + C_{22}(\xi_1,\xi_2)}{C_{11}(\xi_1,\xi_2)}}.
\end{equation*}

The proof of the other part of the theorem consists of 4 steps. 
By lemma \ref{lem:1} we may assume that the matrix $C(\alpha)$ is identity, where$\alpha = (\alpha_1, \dots, \alpha_n)$. Let us introduce the denotation:
$$
C_{\varphi} = \begin{pmatrix}
	\cos^2(\varphi) & 0 \\
	0          & \sin^2(\varphi) 
\end{pmatrix}
$$

\textit{Step 1.} Let us prove that
$f(\frac{\alpha_1 + \alpha_2}{\sqrt{2}}) = f(\frac{\alpha_1 - \alpha_2}{\sqrt{2}})$. Let us apply the formula \ref{eq:theorem_1} for
$\xi_1 = \frac{\alpha_1}{\sqrt{2}}, \ \xi_2 = \frac{\alpha_2}{\sqrt{2}}$ and for
$\xi_1 = \frac{\alpha_1}{\sqrt{2}}, \ \xi_2 = -\frac{\alpha_2}{\sqrt{2}}$, taking into account that function $f$ is absolutely homogeneous:
$$
f\left(\frac{\alpha_1 + \alpha_2}{\sqrt{2}}\right) = 
F\left(C_{\frac {\pi}4}, \frac{f(\alpha_1)}{\sqrt{2}}, \frac{f(\alpha_2)}{\sqrt{2}}\right)=
f\left(\frac{\alpha_1 - \alpha_2}{\sqrt{2}}\right).
$$

Let us consider a random vector
$\bar{\alpha} = (\bar{\alpha}_1, \dots, \bar{\alpha}_n)$, where
$$
\bar{\alpha}_1 = \frac{\alpha_1 + \alpha_2}{\sqrt{2}}, \ \bar{\alpha}_2 = \frac{\alpha_1 - \alpha_2}{\sqrt{2}}, \ \bar{\alpha}_i = \alpha_i \quad \text{for all} \quad i = 3, \dots, n.
$$ 
Easy to see that $\bar{\alpha}$ is an admissible vector, $V(\alpha) = V(\bar{\alpha})$ and $C(\bar{\alpha}) = \mathrm{E}$. Without loss of generality we can assume that the vector $\bar{\alpha}$ was chosen as the initial value of the vector $\alpha$. Then we have $f(\alpha_1) = f(\alpha_2)$.

\medskip
\textit{Step 2.} Let us define a function $T:\mathbb{R} \to \mathbb{R}$ by $T(\varphi) = f(\cos(\varphi)\alpha_1 + \sin(\varphi)\alpha_2)$ for all $\varphi \in \mathbb{R}$.
Let us prove that
\begin{equation} \label{eq:z=z+pi/2}
	T(\varphi) = T\left(\varphi + \frac{\pi}2\right) \quad \text{for all} \quad \varphi \in \mathbb{R}.
\end{equation}
Let us apply the formula \ref{eq:theorem_1} for
$\xi_1 = \cos(\varphi)\alpha_1, \ \xi_2 = \sin(\varphi)\alpha_2$ and for $\xi_1 = \cos(\varphi)\alpha_2, \ \xi_2 = -\sin(\varphi)\alpha_1$, taking into account $f(\alpha_1) = f(\alpha_2)$ and the fact that the function $f$ is absolutely homogeneously:
\begin{multline*}
	T(\varphi) = f(\cos(\varphi)\alpha_1 + \sin(\varphi)\alpha_2) = \\
	= F(C_{\varphi}, |\cos(\varphi)|f(\alpha_1), |\sin(\varphi)|f(\alpha_2)) = \\
	= F(C_{\varphi}, |\cos(\varphi)|f(\alpha_2), |\sin(\varphi)|f(\alpha_1)) = \\ = f(\cos(\varphi)\alpha_2 - \sin(\varphi)\alpha_1) =  T\left(\varphi + \frac{\pi}2\right).
\end{multline*}

\textit{Step 3.} Let us take an arbitrary number $z \in \mathbb{R}$ and prove that
\begin{equation} \label{eq:z+phi}
	T(z + \varphi) = T\left(z - \varphi\right) \quad \text{for all} \quad \varphi \in \mathbb{R}.
\end{equation}
Let us define random variables
$$
\beta_1 = \cos(z)\alpha_1 + \sin(z) \alpha_2, \quad \beta_2 = -\sin(z)\alpha_1 + \cos(z) \alpha_2.
$$
From (\ref{eq:z=z+pi/2}) it follows that $f(\beta_1) = T(z) = T\left(z + \frac{\pi}2\right) = f(\beta_2)$. Then
\begin{multline*}
	T(z \pm \varphi) = f(\cos(z + \varphi)\alpha_1 + \sin(z \pm \varphi)\alpha_2) = 
	f(\cos(\varphi)\beta_1 \pm \sin(\varphi)\beta_2)= \\
	= F(C_{\varphi}, |\cos(\varphi)|f(\beta_1), |\sin(\varphi)|f(\beta_2)). 
\end{multline*}
Substituting $\varphi = z$ in (\ref{eq:z+phi}) yields $T(0) = T(2z)$. This means that the function $T(z)$ is constant and  on the two-dimensional space of linear combinations $\alpha_1$ and $\alpha_2$ the function $f(\xi)$ is equal to $f_0\sqrt{\mathrm{D}\xi}$.

\medskip
\textit{Step 4.} Now we are ready to complete the proof of the theorem. Let us prove that for any random variables
$\xi_1 \in V(\alpha),\  \xi_2 \in V(\alpha)$ it holds that $\frac{f(\xi_1)}{\sqrt{\mathrm{D}\xi_1}} = \frac{f(\xi_2)}{\sqrt{\mathrm{D}\xi_2}}$. Without loss of generality we may assume that $\mathrm{D}\xi_1 = \mathrm{D}\xi_2 = 1$ and that $\xi_1$ is linear independent with $\xi_2$. Let us define
$$
\gamma_1 = \frac{\xi_1 + \xi_2}{\sqrt{\mathrm{D}(\xi_1 + \xi_2)}}, \quad  
\gamma_2 = \frac{\xi_1 - \xi_2}{\sqrt{\mathrm{D}(\xi_1 - \xi_2)}}.
$$
Then $\mathrm{D}\gamma_1 = \mathrm{D}\gamma_2 = 1$ and $\mathrm{cov}(\gamma_1, \gamma_2) = 0$. We extend $\gamma_1$ and $\gamma_2$ to an admissible random vector $\gamma = (\gamma_1, \dots, \gamma_n)$, such that $V(\gamma) = V(\alpha)$ and $C(\gamma) = \mathrm{E}$. Let us repeat the steps 1 --- 3 for the vector $\gamma$ instead the vector $\alpha$. We obtain that on the two dimensional space of linear combinations $\gamma_1$ and $\gamma_2$ the function $f(\xi) = f_0\sqrt{\mathrm{D}\xi}$, i.e. $f(\xi_1) = f_0 = f(\xi_2)$. $\hfill \square$


\section{Numerical experiments.} \label{sec:experiments}

\noindent
\textit{Aim.} We want to check the accuracy of the approximation of portfolio turnover by the estimations (\ref{eq:zura_est_2}), (\ref{eq:zura_est_spectral}) and (\ref{new_est}), depending on the <<spread width>> of the ratio $\frac{\tau}{std}$ for alphas.
    
\medskip
\noindent
\textbf{Procedure.}
    \begin{itemize}
    \item We start with constructing several alphas and select 3 sets of them with big, medium and small <<spread width>> of the ratio $\frac{\tau}{std}$ correspondingly. 
    \item Then we suggest <<metrics>> for portfolio turnover estimations to measure the accuracy of the real turnover approximation.
    \item For each pair of alphas from a set we calculate <<metrics>> for turnover estimations for all two--alphas portfolios and average it by all pairs of alphas.
    \item We calculate averaged <<metrics>> for turnover estimations for 100 portfolios consisting of all alphas of a set with weights which are proportional to coordinates of Sobol pseudorandom sequence of points.
            
\end{itemize}


\subsection{Building and examples of alphas.}

    We use daily data \textit{open, high, low, close, volume} from Yahoo Finance for approximately 1400 most liquid USA stocks. For alpha research we use 2010 --- 2014 as in-sample, and use 02.01.2018 -- 14.06.2024 for testing  as out-of-sample. Below are some alpha examples (without operations):
	
	\begin{enumerate}		 
		\item $\frac{sum(volume,4)\sqrt{high*low}}{sum(close*volume,4)} - 1,$
		\item $\left(\frac{delay(close,14)}{close} - 1\right)\left(\frac{volume}{sum(volume,30)}\right)$,
		\item $correlation\left(close, volume, 20\right)\left(1-\frac{delay(close,10)}{close}\right)$,
		\item $-rsi(close,14)$, where $rsi$ --- is the relative strength index.
	\end{enumerate}
	
	\medskip
	For research we use also alpha operations: truncate, decay, cutting extremes or middles, neutralization and normalization (the latter two are obligatory).


	\subsection{Alpha numerical characteristics.}
        \label{subsec:alpha_stats}
	
    For the given dataset with daily data let us enumerate days from 1 to $p$ in chronological order, i.e. the eldest day is 1. Let's enumerate the stocks with numbers from 1 to $s$.

    \noindent
    Alpha \textbf{position vector} at the day $d$ for $d = 1, \dots, p$ is the following: 
    $$
    a(d) = (a_1(d), \dots, a_s(d))^{\top}.
    $$
	
    \noindent
    \textbf{Return} of the $i$-th stock at the day $d$ for $i = 1, \dots, s$ and $d = 2, \dots, p$: 
	$$
	return_i(d) = \frac{close_i(d)}{close_i(d-1)}-1.
	$$
	
	\noindent
    \textbf{Alpha PnL\footnote{It is holding PnL for delay 1 alphas, see the book \cite{Tulch} for details. These alphas calculate position vectors by the beginning of the trading day, so the execution has the whole day to achieve the required positions.  This provide the possibility to increase the size of investments.}} at the day $d$ for $d = 2, \dots, p$: 
	\begin{equation} \label{eq:holding_pnl}
		PnL(d) =  \displaystyle\sum_{i=1} ^{s} return_i(d)a_i(d-1).
	\end{equation} 
	
	\noindent
	\textbf{Cumulative PnL} by the day $k$ for $k = 2, \dots, p$: 
	$$
	cumPnL(k) =  \displaystyle\sum_{d=2} ^{k} PnL(d).
	$$
	
	\noindent	
	\textbf{Volatility} (sample standard deviation of returns vector): 
	$$
	stdPnL(k) = \sqrt{\frac{1}{k - 2} \displaystyle\sum_{d = 2}^{k} \left(PnL(d) - \frac{1}{k-1}cumPnL(k)\right)^2}
	$$
	
	\noindent
	\textbf{Sharpe ratio\footnote{
			For non dollar neutral portfolios Sharpe ratio use the risk-free rate $R_f$: $\mathrm{sharpe} = \sqrt{T}\frac{\mathrm{meanPnl}(k) - R_f}{\mathrm{stdPnl}(k)}$.} } (252 is a number of trading days in a year): 
	$$
	sharpe = \frac{\sqrt{252}}{p-1}\frac{cumPnL(p)}{stdPnL(p)}.
	$$
	
    \noindent
    Alpha \textbf{turnover} at the day $d$ for $d = 2, \dots, p$: 
    $$
    \tau(d) = \displaystyle\sum_{i=1} ^{s} |a_{i}(d) - a_{i}(d-1)|.
    $$
    Kakushadze in \cite{Zura_10} and Kakushadze with Tulchinsky in \cite{Zura_12} investigated numerical characteristics of alphas, which are real traded on USA stock market. Regression analysis shows that the PnL essentially depends on the alpha returns volatility, return by share depends on the Sharpe ratio and position holding period, while PnL and pairwise correlations weakly depend on the turnover.
    

	\subsection{<<Metrics>> for turnover estimations.}
		
	Let $T_{\ast}(d)$, $\tau(d)$ and $T_{max}(d)$ be the portfolio turnover estimation, real portfolio turnover (i.e. with crossing of trades) and turnover of the portfolio without crossing of trades at the day $d$ correspondingly. Let $p$ be the number of trading days in the testing dataset. Let us define the following <<metrics>>:

	\begin{multline} \label{metrics}
		\hfill
		\rho_{1}(T_{*}) = \frac{1}{p}\displaystyle\sum_{d=1} ^{p} (T_{*}(d) - \tau(d)),
		\hfill
		\rho_{2}(T_{*}) = \frac{1}{p} \displaystyle\sum_{d = 1}^{p} |T_{*}(d) - \tau(d)|, 
		\hfill \\ \hfill
		\rho_{3}(T_{*}) =  \frac{\rho_{1}(T_{*})}{\rho_{1}(T_{max})}, 
		\hfill
		\rho_{4}(T_{*}) =  \frac{\rho_{2}(T_{*})}{\rho_{1}(T_{max})},
		\hfill
		\rho_{5}(T_{*}) = \displaystyle \sum_{d=1}^{p} \frac{|T_{*}(d) - \tau(d)|}{\tau(d)}.
		\hfill
	\end{multline}

	Note that $\rho_{2}$ and $\rho_{5}$ are the absolute and the relative errors correspondingly. Also $\rho_{1}(T_{max}) > 0.$


	\subsection{Experiment results.}
	
\begin{figure}[h]
	\centering  
	\includegraphics[width=0.7\textwidth]{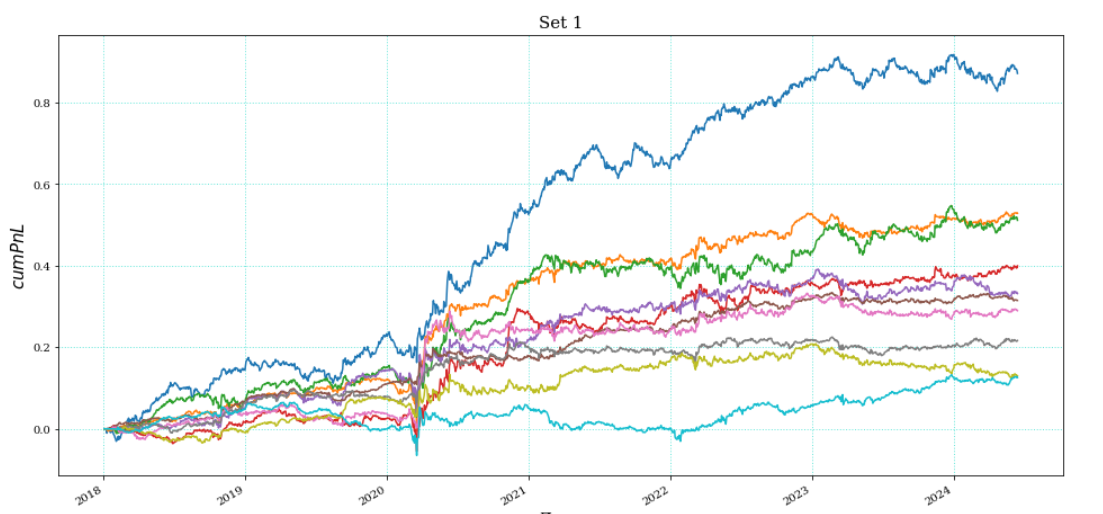}
        \caption{cumulative PnL of alphas in the set 1.}
        \label{pic:1}	
\end{figure}

\begin{figure}[h]
	\centering  
	\includegraphics[width=0.7\textwidth]{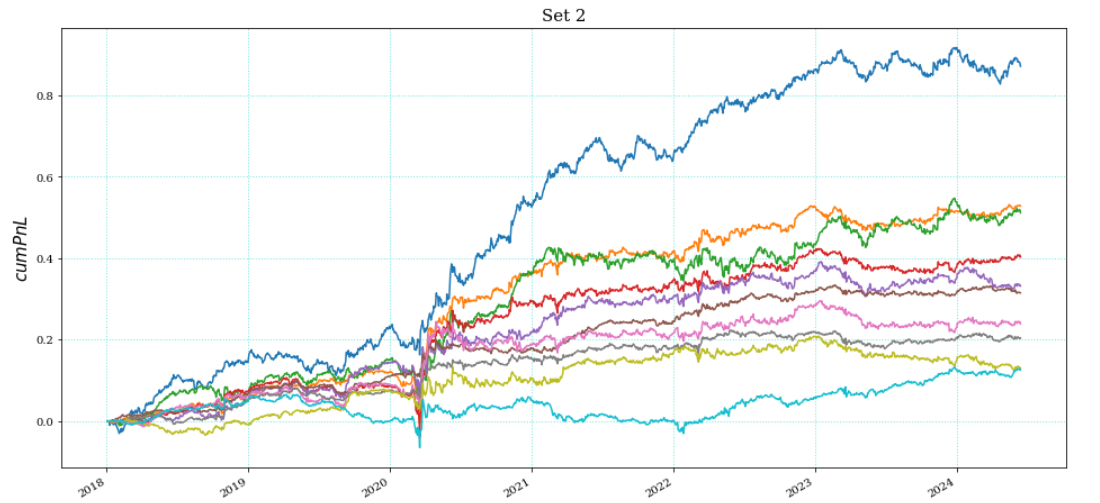}
        \caption{cumulative PnL of alphas in the set 2.}
        \label{pic:2}	
\end{figure}

\begin{figure}[h]
	\centering  
	\includegraphics[width=0.7\textwidth]{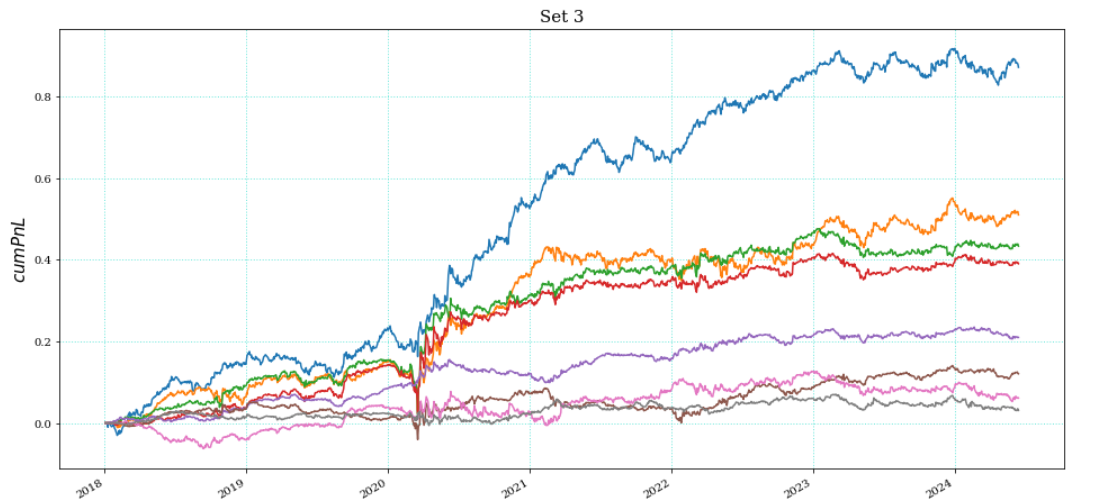}
        \caption{cumulative PnL of alphas in the set 3.}
        \label{pic:3}	
\end{figure}

    Alpha sets 1 -- 3 are developed to produce low pairwise correlations of returns and appropriate returns. Alpha can be included in several sets: so sets 1 and 2 have alphas 2, 3, 5, 6, 9, 10 in common, and alpha 1 is in all three sets. Alpha characteristics from the table \ref{table:stats} (cumulative PnL, Sharpe ratio, standard deviation of returns, daily averaged turnover and the ratio of the daily averaged turnover to standard deviation of returns) and pairwise correlations from the table \ref{table:corr} are calculated during the testing period 02.01.2018 -- 14.06.2024. Alpha cumulative PnLs are shown on figures \ref{pic:1} --- \ref{pic:3}. It happens that the ratio $\frac{\tau}{stdPnL}$ differs up to 7 times, up to 4 times and up to 24\% for alphas in 1, 2 and 3 set correspondingly. 

    For tables \ref{table:1} and \ref{table:2} the mean alpha turnover and covariance matrix of returns for estimations $T_{*1} - T_{*4}$ are calculated in rolling window of 250 days. For estimation $T_{KL}$ covariance matrix of returns is calculated using all testing period, then we find its eigenvectors and eigenvalues which are used in the estimation. 
    The table \ref{table:1} is built as follows. Each pair of alphas form a portfolio with weights $x_1 = x_2 = \frac{1}{2}$. Then we calculate turnover estimations and the real turnover, <<metrics>>, take the absolute value for $\rho_1$ and $\rho_3$, and average by all pairs of alphas in a set. In turn, the table \ref{table:2} is built as follows. For every set of alphas we take 100 portfolios of $n$ alphas, so that weights of the $k$-th portfolio are proportional to the coordinates of the $k$-th point of Sobol pseudorandom sequence in $\mathbb{R}^n$, where $n = 10$ for sets 1 and 2 and $n = 8$ for set 3, also $k = 1,\dots, 100$. For each portfolio we calculate turnover estimations and the real turnover. Then we take absolute value of <<metrics>> and average it by 100 portfolios of the corresponding set. Estimations $T_{*1}$ -- $T_{*4}$ are the same as in \ref{metrics}; estimation $T_{KL}$ is given by \ref{eq:zura_est_2} and \ref{eq:zura_est_spectral} for tables \ref{table:1} and \ref{table:2} correspondingly; estimation $T_{max}$ is a turnover without crossing of trades, i.e. is the linear combination of alphas turnover.

    The estimation is better, if the <<metric>> is less. Tables results allow us to make the following observations.
    \begin{itemize}
	\item <<Metrics>> $\rho_{1}$ and $\rho_{2}$ for sets 1 and 2 differ a little. This mean that estimations are typically (for most of days) shifted in one direction.
	
	\item For all estimations their absolute error $\rho_{2}$ and also $\rho_{1}$ decrease while moving from set 1 to set 3 through set 2.
	\item The relative error $\rho_5$, $\rho_{3}$ and $\rho_{4}$ stay approximately the same for estimation $T_{KL}$ and decrease for estimations $T_{*1}-T_{*4}$ while moving from set 1 to set 3. 
	
	\item Estimations $T_{*1}-T_{*4}$ outperform the estimation $T_{KL}$ for the set 3. 
	\end{itemize}
    
    \begin{table}[h!]
    \centering
	\resizebox{\textwidth}{!}{\begin{tabular}{|l|l|l|l|l|l|l|l|l|l|l|l|l|l|l|l|}
		\hline
            & \multicolumn{5}{|c|}{Set 1} & \multicolumn{5}{|c|}{Set 2} & \multicolumn{5}{|c|}{Set 3} \\
        \hline
		  & $\rho_{1}$ & $\rho_{2}$ & $\rho_{3}$ & $\rho_{4}$ & $\rho_{5}$ & $\rho_{1}$ & $\rho_{2}$ & $\rho_{3}$ & $\rho_{4}$ & $\rho_{5}$ & $\rho_{1}$ & $\rho_{2}$ & $\rho_{3}$ & $\rho_{4}$ & $\rho_{5}$\\ 
		\hline
		$T_{KL}$  & 0.023 & 0.028 & 0.527 & 0.597 & 0.089 & 0.023 & 0.027 & 0.561 & 0.628 & 0.108 & 0.013 &	0.016 & 0.528 & 0.628 & 0.125 \\ 
		$T_{*1}$  & 0.039 & 0.044 & 0.613 & 0.720 & 0.122 & 0.030 & 0.034 & 0.567 & 0.674 & 0.138 & 0.007 & 0.012 & 0.278 & 0.458 & 0.095 \\ 
		$T_{*2}$  & 0.058 & 0.062 & 1.092 & 1.150 & 0.152 & 0.025 & 0.029 & 0.570 & 0.662 & 0.112 & 0.008 & 0.012 & 0.309 & 0.469 & 0.093 \\
		$T_{*3}$  & 0.039 & 0.044 & 0.613 & 0.720 & 0.122 & 0.030 & 0.034 & 0.567 & 0.674 & 0.138 & 0.007 & 0.012 & 0.278 & 0.458 & 0.095 \\
		$T_{*4}$  & 0.029 & 0.033 & 0.542 & 0.603 & 0.087 & 0.023 & 0.027 & 0.493 & 0.563 & 0.102 & 0.007 & 0.011 & 0.280 & 0.409 & 0.084 \\
		$T_{max}$ & 0.067 & 0.067 & 1.000 & 1.000 & 0.185 & 0.052 & 0.052 & 1.000 & 1.000 & 0.211 & 0.027 & 0.027 & 1.000 & 1.000 & 0.215 \\ 
		\hline		
	\end{tabular}
    }
	\caption{Averaged by all alpha pairs <<metrics>> of turnover estimations.}
	\label{table:1}	
	\end{table}	
    
    \begin{table}[h!] 
    \centering
    \resizebox{\textwidth}{!}{\begin{tabular}{|l|l|l|l|l|l|l|l|l|l|l|l|l|l|l|l|}
    \hline
        & \multicolumn{5}{|c|}{Set 1} & \multicolumn{5}{|c|}{Set 2} & \multicolumn{5}{|c|}{Set 3} \\
    \hline
        & $\rho_{1}$ & $\rho_{2}$ & $\rho_{3}$ & $\rho_{4}$ & $\rho_{5}$ & $\rho_{1}$ & $\rho_{2}$ & $\rho_{3}$ & $\rho_{4}$ & $\rho_{5}$ & $\rho_{1}$ & $\rho_{2}$ & $\rho_{3}$ & $\rho_{4}$ & $\rho_{5}$\\ 
    \hline
        $T_{KL}$  & 0.112 & 0.112 & 0.677 & 0.677 & 0.405 & 0.076 & 0.076 & 0.644 & 0.644 & 0.401 & 0.033 & 0.033 & 0.604 & 0.604 & 0.328 \\ 
        $T_{*1}$  & 0.054 & 0.061 & 0.355 & 0.394 & 0.243 & 0.059 & 0.059 & 0.508 & 0.510 & 0.326 & 0.009 & 0.015 & 0.157 & 0.276 & 0.152 \\ 
        $T_{*2}$  & 0.036 & 0.043 & 0.223 & 0.269 & 0.153 & 0.031 & 0.034 & 0.269 & 0.292 & 0.189 & 0.007 & 0.014 & 0.118 & 0.255 & 0.138 \\
        $T_{*3}$  & 0.052 & 0.055 & 0.321 & 0.336 & 0.210 & 0.058 & 0.058 & 0.491 & 0.491 & 0.314 & 0.009 & 0.014 & 0.154 & 0.260 & 0.143 \\
        $T_{*4}$  & 0.040 & 0.046 & 0.261 & 0.297 & 0.182 & 0.039 & 0.041 & 0.342 & 0.354 & 0.228 & 0.007 & 0.014 & 0.132 & 0.253 & 0.138 \\
        $T_{max}$ & 0.161 & 0.161 & 1.000 & 1.000 & 0.591 & 0.117 & 0.117 & 1.000 & 1.000 & 0.624 &  0.055 & 0.055 & 1.000 & 1.000 & 0.540 \\ 
		\hline		
	\end{tabular}
    }
	\caption{<<Metrics>> for turnover estimations, a portfolio consists of all alphas of a set.}
	\label{table:2}	
	\end{table}	


	\section{Conclusions.} \label{sec:conclusions}
	We investigate the approximations of an $n$-alpha portfolio turnover via covariance matrix of alpha returns, alpha turnovers and weights, provided crossing of trades of opposite signs. Estimations become more precise in the case when the ratios $\frac{\tau_i}{std_i}$ differ a little for alphas of portfolio. In this case proposed estimations (\ref{new_est}) seems to be better choice. Else, if ratio $\frac{\tau_i}{std_i}$ differ significantly, then estimations (\ref{eq:zura_est_2}) or (\ref{eq:zura_est_spectral}) are preferred.
	
    \textbf{Practical value of the work.}
    \textit{Portfolio managers got an understanding that turnover estimations may be more precise and the possibility to achieve it. One need to make the ratios $\frac{\tau_i}{std_i}$ closer to each other.}

    More precise turnover estimations produce more adequate optimized function for portfolio construction via optimization methods, provided accounting transaction costs. This may lead to <<better>> portfolio.
	
	\textbf{Theoretical value of the work.}
	\textit{The theorem \ref{th:main} could help to analyse the models of other absolutely homogeneous functions of portfolio, which depend on covariance matrix of returns.}


\newpage


\begin{table}[h]
    \begin{tabular}{|l|l|l|l|l|l|l|l|l|l|l|}
    \hline
    Set 1 & alpha 1 & alpha 2 & alpha 3 & alpha 4 & alpha 5 & alpha 6 & alpha 7 & alpha 8 & alpha 9 & alpha 10 \\ 
    \hline			
    $ cumPnL $ & 0.870 & 0.530 & 0.510 & 0.400 & 0.330 & 0.310 & 0.290 & 0.220 & 0.130 &  0.130 \\
    $ Sharpe $ & 1.482 & 1.179 & 0.932 & 0.861 & 0.692 & 1.344 & 0.704 & 0.795 & 0.369 & 0.385 \\
    $ T $ & 0.217 & 0.571 & 0.297 & 0.963 & 0.176 & 0.213 & 0.767 &  0.735 & 0.223 & 0.216 \\
    $ stdPnL $ & 0.0057 & 0.0044 & 0.0054 & 0.0045 & 0.0047 & 0.0023 & 0.004 & 0.0027 & 0.0034 & 0.0032 \\
    $ T/stdPnL $ & 37.69 & 130.29 & 55.42 & 212.30 & 37.48 & 92.87 & 190.77 & 276.31 & 64.96 & 67.57 \\	
    \hline \hline
    
    Set 2 & alpha 1 & alpha 2 & alpha 3 & alpha 4 & alpha 5 & alpha 6 & alpha 7 & alpha 8 & alpha 9 & alpha 10 \\ 
    \hline
    $ cumPnL $ & 0.870 & 0.530 & 0.510 & 0.400 & 0.330 & 0.310 & 0.240 & 0.200 & 0.130 &  0.130 \\
    $ Sharpe $ & 1.482 & 1.179 & 0.932 & 0.857 & 0.692 & 1.344 & 0.598 & 0.816 & 0.369 & 0.385 \\
    $ T $ & 0.217 & 0.571 & 0.297 & 0.355 & 0.176 & 0.213 & 0.430 &  0.383 & 0.223 & 0.216 \\
    $ stdPnL $ & 0.0057 & 0.0044 & 0.0054 & 0.0046 & 0.0047 & 0.0023 & 0.0039 & 0.0024 & 0.0034 & 0.0032 \\
    $ T/stdPnL $ & 37.69 & 130.29 & 55.42 & 77.20 & 37.48 & 92.87 & 110.16 & 157.02 & 64.96 & 67.57 \\
    \hline 
    \end{tabular}
    
    \begin{tabular}{|l|l|l|l|l|l|l|l|l|}
    \hline
    Set 3 & alpha 1 & alpha 2 & alpha 3 & alpha 4 & alpha 5 & alpha 6 & alpha 7 & alpha 8 \\ 
    \hline
    $ cumPnL $ & 0.870 & 0.510 & 0.430 & 0.390 & 0.210 & 0.120 & 0.060 & 0.030 \\
    $ Sharpe $ & 1.482 & 0.911 & 0.890 & 0.857 & 0.920 & 0.395 & 0.169 & 0.153 \\       
    $ T $ & 0.217 & 0.231 & 0.209 & 0.163 & 0.098 & 0.119 & 0.152 &  0.090 \\
    $ stdPnL $ & 0.0057 & 0.0055 & 0.0048 & 0.0044 & 0.0022 & 0.003 & 0.0035 & 0.002 \\
    $ T/stdPnL $ & 37.69 & 42.20 & 43.83 & 36.60 & 43.94 & 39.77 & 42.93 & 45.35 \\
    \hline                 
    \end{tabular}
    \caption{Alpha characteristics.}
    \label{table:stats}	
\end{table}

\begin{table}[h]
    \begin{tabular}{ |c|c|c|c|c|c|c|c|c|c|}
    \hline
    Set 1 & alpha 2 & alpha 3 & alpha 4 & alpha 5 & alpha 6 & alpha 7 & alpha 8 & alpha 9 & alpha 10 \\ 
    \hline			
    alpha 1  & 0.27 & 0.61 & 0.19 & 0.24 & 0.13 & 0.13 & 0.10 & 0.05 & 0.24 \\  
    alpha 2 & 1.00 & 0.60 & 0.65 & 0.74 & 0.56 & 0.65 & 0.62 & 0.59 & 0.46 \\
    alpha 3 & & 1.00 & 0.38 & 0.52 & 0.36 & 0.35 & 0.27 & 0.19 & 0.51 \\   
    alpha 4 & & & 1.00 & 0.32 & 0.27 & 0.42 & 0.59 & 0.33 & 0.22 \\ 
    alpha 5 & & & & 1.00 & 0.73 & 0.55 & 0.47 & 0.72 & 0.46 \\  
    alpha 6 & & & & & 1.00 & 0.49 & 0.52 & 0.62 & 0.34 \\       
    alpha 7 & & & & & & 1.00 & 0.52 & 0.51 & 0.28 \\  
    alpha 8 & & & & & & & 1.00 & 0.47 & 0.2 \\           
    alpha 9 & & & & & & & &  1.00 & 0.05 \\             
    \hline \hline
    
    Set 2 &  alpha 2 & alpha 3 & alpha 4 & alpha 5 & alpha 6 & alpha 7 & alpha 8 & alpha 9 & alpha 10 \\ 
    \hline			
    alpha 1 & 0.27 & 0.61 & 0.24 & 0.24 & 0.13 & 0.16 & 0.08 & 0.05 & 0.24 \\                 
    alpha 2 & 1.00 & 0.60 & 0.85 & 0.74 & 0.56 & 0.64 & 0.66 & 0.59 & 0.46 \\
    alpha 3 & & 1.00 & 0.54 & 0.52 & 0.36 & 0.40 & 0.28 & 0.19 & 0.51 \\  
    alpha 4 & & & 1.00 & 0.79 & 0.61 & 0.80 & 0.70 & 0.68 & 0.44 \\  
    alpha 5 & & & & 1.00 & 0.73 & 0.79 & 0.68 & 0.72 & 0.46 \\   
    alpha 6 & & & & & 1.00 & 0.62 & 0.71 & 0.62 & 0.34 \\        
    alpha 7 & & & & & & 1.00 & 0.63 & 0.64 & 0.34 \\           
    alpha 8 & & & & & & & 1.00 & 0.64 & 0.27 \\           
    alpha 9 & & & & & & & &  1.00 & 0.05 \\             
    \hline                
    \end{tabular}

    \begin{tabular}{ |c|c|c|c|c|c|c|c|c|}
    \hline
    Set 3 & alpha 2 & alpha 3 & alpha 4 & alpha 5 & alpha 6 & alpha 7 & alpha 8 \\ 
    \hline			
    alpha 1 & 0.62 & 0.24 & 0.33 & 0.14 & 0.30 & 0.06 & 0.42 \\      
    alpha 2 & 1.00 & 0.52 & 0.62 & 0.28 & 0.63 & 0.09 & 0.40 \\
    alpha 3 & & 1.00 & 0.90 & 0.49 & 0.49 & 0.56 & 0.08 \\  
    alpha 4 & & & 1.00 & 0.59 & 0.50 & 0.67 & 0.19 \\  
    alpha 5 & & & & 1.00 & 0.24 & 0.65 & -0.03 \\       
    alpha 6 & & & & & 1.00 & -0.03 & 0.03 \\           
    alpha 7 & & & & & & 1.00 & 0.10 \\           
    \hline
    \end{tabular}
    \caption{Pairwise correlations of alpha returns.}
    \label{table:corr}	
\end{table}


\newpage

\begin{thebibliography}{99}

\bibitem{Chan} E.P. Chan. Algorithmic trading. Wiley, 2013.

\bibitem{Tulch} I. Tulchinsky et al. Finding Alphas: A Quantitative Approach to Building Trading Strategies. New York, NY: Wiley, 2020.

\bibitem{Zura_11} Z. Kakushadze, J.A. Serur. 151 Trading Strategies. Cham, Switzerland: Palgrave Macmillan, an imprint of Springer Nature, 1st Edition, XX, 480 pp., 2018. 


\bibitem{Zura_10} Z. Kakushadze. 101 Formulaic Alphas. Wilmott Magazine, 84: 72--80, (2016).

\bibitem{Zura_1} Z. Kakushadze, J.K.-S. Liew. Is it possible to od on alpha? 
The Journal of Alternative Investments, 18(2): 39--49, (2015).

\bibitem{Zura_2} Z. Kakushadze. Spectral model of turnover reduction. 
Econometrics, 3(3): 577--589, (2015).

\bibitem{Zura_4} Z. Kakushadze. Can Turnover Go to Zero? 
Journal of Derivatives \& Hedge Funds, 20(3): 157--176, (2014). 



\bibitem{Markowitz} H. Markowitz. Portfolio selection. The Journal of Finance, 7(1): 77--91, (1952).


\bibitem{Bruder} Bruder, B. and Roncalli, T. (2012). Managing risk exposures using the risk budgeting approach.	Available at SSRN 2009778.

\bibitem{Zura_9} Z. Kakushadze, W. Yu. Notes on Fano Ratio and Portfolio Optimization.	
Journal of Risk \& Control, 5(1): 1--33, (2018). 



\bibitem{Zura_3} Z. Kakushadze. Combining Alpha Streams with Costs. 
Journal of Risk, 17(3): 57--78, (2015). 


\bibitem{Zura_5} Z. Kakushadze. Notes on Alpha Stream Optimization.
Journal of Investment Strategies, 4(3): 37--81, (2015). 


\bibitem{Zura_6} Z. Kakushadze. Factor Models for Alpha Streams. 
Journal of Investment Strategies, 4(1): 83--109, (2014). 

\bibitem{shrink_1} O. Ledoit, M. Wolf. “Honey, I Shrunk the Sample Covariance Matrix.” The
Journal of Portfolio Management, 30(4): 110--119, (2004).	



\bibitem{Zura_12} Z. Kakushadze, I. Tulchinsky. Performance v. Turnover: A Story by 4,000 Alphas. The Journal of Investment Strategies, 5(2): 75--89, (2016).



\end{thebibliography}
\end{document}